\documentclass{ws-procs9x6}

\usepackage{graphicx}

\newcommand{\beq}{\begin{equation}}
\newcommand{\eeq}{\end{equation}}
\newcommand{\beqa}{\begin{eqnarray}}
\newcommand{\eeqa}{\end{eqnarray}}
\newcommand{\ben}{\begin{enumerate}}
\newcommand{\een}{\end{enumerate}}
\newcommand{\bit}{\begin{itemize}}
\newcommand{\eit}{\end{itemize}}
\newcommand{\rosq}{\langle r^2 \rangle_D^s}

\newcommand{\rmsq}{\langle r^2 \rangle_M^s}

\def\FOS{{F_1^{s}}}
\def\FTS{{F_2^{s}}}

\def\GES{{G_{E}^{s}}}
\def\GMS{{G_{M}^{s}}}
\def\etal{{{\em   et al.}}}
\def\bra#1{{\langle#1\vert}}
\def\ket#1{{\vert#1\rangle}}
\def\pbar{{\bar{p}}}

\def\mn{{M}}
\def\mns{{M^2}}
\def\mks{{m_{K}^2}}

\def\fks{{F_{K}^s}}

\def\mpis{{m_\pi^2}}
\def\bpp{{b_1^{1/2,\,1/2}}}
\def\bpm{{b_1^{1/2,\,-1/2}}}
\def\bppm{{b_1^{1/2,\,\pm 1/2}}}

\begin{document}

\title{NUCLEON STRANGENESS IN DISPERSION THEORY}

\author{H.-W. HAMMER}

\address{Department of Physics\\
The Ohio State University\\
Columbus, OH 43210, USA\\[0.3cm] 
and
}

\address{Helmholtz-Institut f\"ur Strahlen-
und Kernphysik (Abt. Theorie)\\
Universit\"at Bonn\\
Nussallee 14-16, 53115 Bonn, Germany\\
E-mail: hammer@itkp.uni-bonn.de}


\maketitle

\abstracts{
Dispersion relations provide a powerful tool to describe the
low-energy structure of hadrons.
We review the status of the strange vector form 
factors of the nucleon in dispersion theory.
We also comment on open questions and the relation to chiral
perturbation theory.
}

\section{Introduction}
\label{sec:intro}
The low-energy structure of the nucleon's $\bar{s}s$ sea is an
important topic in hadron physics\cite{Mus94a,BeH01}. 
While deep inelastic scattering has provided information about 
the light-cone momentum distribution of the strange sea\cite{CCF93}, 
little is known about the corresponding spatial and spin distributions 
or about the role played by the sea in the nucleon's response to a 
low-energy probe.  Parity-violating experiments
with polarized electrons can be used to probe nucleon
matrix elements of the strange-quark vector current, which is parametrized
by the strange electric and magnetic form factors $\GES$ and $\GMS$,
respectively. In an effort to measure these form factors, several such
experiments have been carried out or are under way at MIT-Bates 
(SAMPLE), Jefferson Lab (HAPPEX, G0), and Mainz (A4).  
These experiments have produced the first
measurements of $\GMS$~\cite{sample}, as well as a linear combination 
of $\GMS$ and $\GES$~\cite{happex}.

Since the strange quark mass is of the order of $\Lambda_{QCD}$,
calculating the strange vector form factors is a complex problem
that crucially depends on the nonperturbative aspects of QCD. A variety of
approaches have been used to calculate the leading strange moments
in the past: lattice QCD, chiral perturbation theory,
a plethora of hadronic models, and dispersion relations.
Most calculations agree neither in sign nor in magnitude. This
reflects the difficulty in calculating $SU(3)$ flavor singlet observables
which are not constrained by the mainly electromagnetic data available.
In this talk, I will focus on what has been learned using dispersion
relations (DR's). After a brief review of the formalism in 
Sec.~\ref{sec:form}, I will discuss the dispersion theory calculations
of the strange vector form factors in Sec.~\ref{sec:dispff}. 
Section~\ref{sec:chpt} deals with the connection to chiral perturbation theory
and Sec.~\ref{sec:conclu} contains the conclusions.
For a discussion of the other approaches,
see the talks by Pene, Riska, and Holstein, as well as
the recent review article by Beck and Holstein\cite{BeH01}.

\section{Formalism}
\label{sec:form}
Using Lorentz and gauge invariance, the nucleon matrix element of the 
strange vector current operator $\bar{s}\gamma_\mu s$ 
can be parametrized in terms of two form factors,
\begin{equation}
\langle N(p') | \bar{s}\gamma_\mu s | N(p) \rangle = \bar{u}(p')
\left[ \FOS (t) \gamma_\mu +\frac{i}{2 M} \FTS (t)\sigma_{\mu\nu}
q^\nu \right] u(p)\,,
\end{equation}
where $M$ is the nucleon mass and $t=(p'-p)^2=q^2$ 
the four-momentum transfer. In electron scattering, the form factors
are probed at spacelike $t<0$.
$\FOS$ and $\FTS$ are the Dirac and Pauli form factors, respectively.
$\FOS$ vanishes at $t=0$ since the net strangeness of the nucleon 
is zero, while $\FTS$ is normalized to the anomalous strange 
magnetic moment of the nucleon, $\kappa^s$.
The experimental data are usually given for the Sachs form factors,
which are linear combinations of the Dirac and Pauli form factors:
\begin{eqnarray}
\GES(t) &=& \FOS(t) - \tau \FTS(t) \, , 
\qquad
\GMS(t) = \FOS(t) + \FTS(t) \, , 
\label{sachs}
\end{eqnarray}
where $\tau = -t/(4 M^2)$.
In the Breit frame, $\GES$ and $\GMS$ may be interpreted as
the Fourier transforms of the distribution of strangeness
charge and magnetization, respectively.              
Most theoretical calculations have focused on the leading moments
of the form factors,
\begin{eqnarray}
\kappa^{s} = \FTS(0)\,, &\qquad& \mu^{s} =\GMS(0)\,, \nonumber\\
\langle r^2 \rangle_D^{s} = 6\frac{d\FOS}{dt}
\bigg|_{t=0}\,,&\qquad&
\langle r^2 \rangle_{E/M}^{s} = 6\frac{dG_{E/M}^{s}}{dt}
\bigg|_{t=0}\,.
\label{norm}
\end{eqnarray}
Note also that $\mu^{s}\equiv \kappa^s$ since $\FOS(0)=0$.

\section{Strange Vector Form Factors in Dispersion Theory}
\label{sec:dispff}
\subsection{Spectral Decomposition}
\label{sec:specd}
Based on unitarity and analyticity, dispersion relations (DR's) relate
the real and imaginary parts of the strange vector form factors.
Consider, {\it e.g.}, the Pauli form factor $\FTS$. 
We write down an unsubtracted DR of the form
\begin{equation}
\label{disprel}
\FTS(t) = \frac{1}{\pi} \, \int_{t_0}^\infty \frac{{\rm Im}\, 
\FTS(t')}{t'-t-i\epsilon}\, dt'\, ,
\label{emff:disp} 
\end{equation}
where $t_0=9\mpis$ is the threshold of the lowest cut of $\FTS$.
In the case of $\FOS$, the value of the form factor at $t=0$
is known and a once subtracted DR can be used. 
By performing more subtractions, the sensitivity to the imaginary
part at high momentum transfer can be reduced and the convergence
behavior of the DR improved. 
Even though some information about the large-$t$ behavior is
available from perturbative QCD and Regge fits, the question of whether 
a given DR converges remains an assumption and
can not be answered {\it a priori}. 
The imaginary part or spectral function entering Eq.~(\ref{disprel}) 
can be obtained from a spectral decomposition\cite{disp,MHD97}. 
For this purpose, it is convenient to consider the 
strange vector current matrix element in the timelike region,
\begin{eqnarray}
J_\mu &=& \langle N(p) \overline{N}(\bar{p}) | \bar{s}\gamma_\mu s
| 0 \rangle 
\nonumber \\ &=& \label{eqJ}
\bar{u}(p) \left[ \FOS (t) \gamma_\mu +\frac{i}{2 M} 
\FTS(t)\sigma_{\mu\nu}(p+\bar{p})^\nu \right] v(\bar{p})\,,
\end{eqnarray}
where $p\,,\bar{p}$ are the momenta of the nucleon-antinucleon pair
created by the current $\bar{s}\gamma_\mu s$. The four-momentum transfer
in the timelike region is $t=(p+\bar{p})^2 > 0$. 
Using the LSZ formalism, the imaginary part
of the form factors is obtained by inserting a complete set of
intermediate states\cite{disp,MHD97}
\begin{eqnarray}
\label{spectro}
{\rm Im}\,J_\mu &=& \frac{\pi}{\sqrt{Z}}(2\pi)^{3/2}{\mathcal N}\,\sum_n
 \langle p | \bar{J}_N (0) | n \rangle 
\langle n | \bar{s}\gamma_\mu s | 0 \rangle \,v(\bar{p})
\,\delta^4(p+\bar{p}-p_n)\,,
\end{eqnarray}
where ${\mathcal N}$ is a nucleon spinor normalization factor, $Z$ is
the nucleon wave function renormalization, and $\bar{J}_N (x) =
J^\dagger(x) \gamma_0$ with $J_N(x)$ a nucleon source.
\begin{figure}[ht]
\centerline{\includegraphics*[width=7cm,angle=0]{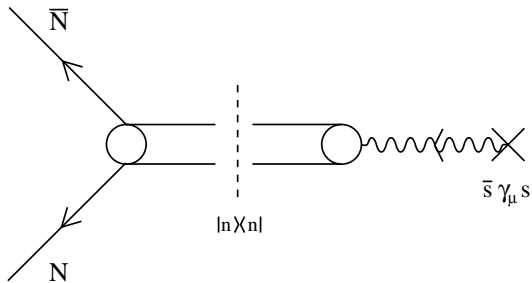}}
\vspace*{-.0in}
\caption{Graphical illustration of the spectral decomposition
in Eq.~(\ref{spectro}). The intermediate states $|n\rangle$ are
on shell.}
\label{fig:specdias}
\end{figure}        
The spectral decomposition in Eq.~(\ref{spectro})
is illustrated in Fig.~\ref{fig:specdias}.
The states $|n\rangle$ are asymptotic states of
momentum $p_n$ and therefore on shell.
They are stable with respect to the strong 
interaction. Only intermediate states 
that carry the same quantum numbers as
the current $\bar{s}\gamma_\mu s$  [$I^G(J^{PC})=0^-(1^{--})$,
baryon number zero, and no net strangeness] contribute to the sum
in Eq.~(\ref{spectro}). The lowest-mass states satisfying this 
criterion are: $3\pi$, $5\pi$, $\ldots$, $K\bar{K}$,
$K\bar{K}\pi$, $K\bar{K}\pi\pi$, $\ldots$, $\Lambda\bar{\Lambda}$, 
$\Sigma\bar{\Sigma}$, $\ldots$.  Because of $G$-parity, only states
with an odd number of pions contribute. 
Associated with each intermediate state is a
cut starting at the corresponding threshold in $t$ and running to
infinity. As a consequence, the spectral 
functions ${\rm Im}\, F_i^s (t),\;i=1,2$ are different from zero along the
cut from $t_0= 9 \, m_\pi^2$ to $\infty$.
Using Eqs.~(\ref{eqJ},\ref{spectro}), the spectral functions 
can in principle be constructed from experimental
data. In practice, this proves a formidable task and has only been
done for the $K\bar{K}$-continuum contribution, as will be discussed
in Sec.~\ref{sec:kkbar}. However, the spectral function 
can also be modelled by using vector meson dominance.

\subsection{Vector Meson Dominance}
\label{sec:vmd}
Within the vector meson dominance (VMD) approach, the spectral
functions are approximated by a few vector meson poles, namely 
the $\omega, \phi, \ldots$.
In that case, the strange vector form factors take the
form
\begin{equation}
F_i^{s} (t) = \sum_{V=\omega,\phi,...} \frac{a_i^{V}}{m^2_{V}-t}
\,, \qquad i = 1,2 \,. 
\label{emff:vmd}
\end{equation}
Clearly, such pole terms contribute to the spectral functions
as $\delta$-functions,
\begin{equation}
{\rm Im}\, F^{s}_i (t) = \pi\, \sum_{V=\omega,\phi,...} a_i^{V} \, 
\delta(t - m_{V}^2) \,, \qquad i = 1,2\, .
\label{emff:imvmd}
\end{equation}
These terms arise naturally as approximations to 
vector meson resonances in the continuum
of intermediate states like $n \pi$ ($n \geq 3)$, 
$K \overline{K}$
and so on. If the continuum contributions are strongly peaked near
the vector meson resonances, Eq.~(\ref{emff:imvmd}) is a good
approximation to the true spectral function.

A dispersion analysis of the nucleon's electromagnetic form factors with
three vector meson poles in both the isoscalar and isovector channels
as well as the two-pion continuum in the isovector channel
was carried out by H{\"o}hler \etal\cite{Hoe76}. 
In this analysis the isoscalar electromagnetic
form factors were parametrized by three vector meson resonances:
$\omega$, $\phi$, and a third fictitious $S'$ meson with a mass of about
$1.6$ GeV. The $S'$ effectively summarizes higher mass strength in
the spectral function.

The first application of this framework  to
the strange vector form factors was carried out by Jaffe\cite{Jaf89}.
He used the knowledge of the flavor content of the $\omega$ and $\phi$
mesons to deduce their coupling to the strange vector current from
their coupling to the isoscalar electromagnetic current
by performing a rotation in flavor space.
The underlying principle is very simple:
The flavor content of the $\omega$ and $\phi$ mesons can be written
as:
\beq
|\phi\rangle \approx |s\bar{s}\rangle\,,\qquad
|\omega\rangle \approx \frac{1}{\sqrt{2}}(|u\bar{u}\rangle 
+|d\bar{d}\rangle)\,,
\eeq
while the isoscalar electromagnetic quark current is simply
\beq
j_\mu^{\rm is}=\frac{1}{6}\left(\bar{u}\gamma_\mu u+\bar{d}\gamma_\mu d\right)
-\frac{1}{3}\bar{s}\gamma_\mu s\,,
\eeq
if the heavier $c$, $b$, and $t$ quarks are neglected.
The OZI rule then leads to the relations
\beq
\langle 0 | \bar{s}\gamma_\mu s |\omega \rangle \approx 0\,,
\qquad
\langle 0 | \bar{u}\gamma_\mu u + \bar{d}\gamma_\mu d|\phi \rangle 
\approx 0\,,
\eeq
which can be used to determine the coupling of the $\omega$ and $\phi$
mesons to the strange vector current. If $\omega-\phi$-mixing and
the phenomenological OZI violation are also included,
one finds\cite{HRM99a}
\beq
\langle 0 | \bar{s}\gamma_\mu s |\phi \rangle \approx
-3 \langle 0 | j_\mu^{\rm is} |\phi \rangle\,,\qquad
\langle 0 | \bar{s}\gamma_\mu s |\omega \rangle \approx
-0.2 \langle 0 | j_\mu^{\rm is} |\omega \rangle\,,
\eeq
which lead to a relation between the pole residues $a_i^V$ for the 
strange vector and isoscalar electromagnetic currents.
The coupling of the fictitiuos $S'$ meson, however, can not be 
deduced this way since its flavor content is unknown. Furthermore,
it is not guaranteed that the strange vector form factor can be described
using the same three poles as for the isoscalar electromagnetic form
factors. Casting this issue aside, one can determine the coupling of the
$S'$ meson by fixing the asymptotic behavior of the form factors.
(But note that this procedure is not unique and leads to an ambiguity in the
leading moments\cite{rm95}.) Using the electromagnetic 
form factor analysis of H{\"o}hler \etal\cite{Hoe76}, Jaffe determined the
the leading moments $\mu^s$ and $\langle r^2 \rangle_D^s$
of the strange form factors\cite{Jaf89}.
In particular, his analysis gave a large strange magnetic
moment. This was due to the strong coupling of the 
$\phi$ to the nucleon in electromagnetic analyses\cite{Hoe76,MMD96},
where a second pole around $t=1.0$ GeV$^2$ 
in addition to the $\omega$ is required 
to generate the dipole behavior of the data. Usually this pole
is identified with the $\phi$, which implies a large OZI violation.

Since the electromagnetic form factor analysis\cite{Hoe76} is over
20 years old and does not include the precise data from later
experiments, it was recently updated and improved\cite{MMD96}. 
Subsequently, we have performed an updated analysis of the strange 
vector form factors based on the new electromagnetic data\cite{HMD96s}.
The results for leading moments are
\beq
\mu^s=-0.24\pm 0.03\,, \qquad 
\langle r^2 \rangle_D^s = (0.21\pm 0.03) \mbox{ fm}^2\,,
\eeq
where the error includes only the uncertainty in the phenomenological
$\phi-\omega$-mixing parameter $\epsilon$.
The systematic error from using the pole ansatz for the spectral function
is difficult to quantify.
\begin{figure}[ht]
\vspace*{-0.2cm}
\centerline{\includegraphics*[width=8cm,angle=0]{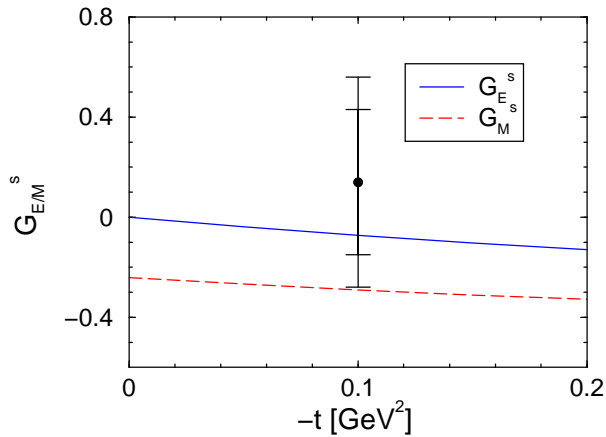}}
\vspace*{-.0in}
\caption{Strange vector form factors from the updated pole
 analysis\protect\cite{HMD96s}. The data point for
 $G_M^s$ is from the SAMPLE collaboration\protect\cite{sample}.
 The inner error bar is statistical, while the outer error bar combines
 statistical and systematic errors in quadrature.}
\label{fig:strange_poles}
\end{figure}        
In Fig.~\ref{fig:strange_poles}, we show the results of this analysis
compared to the data point for $\GMS$ from the SAMPLE 
collaboration\cite{sample}. The calculation breaks down at
higher momentum transfers. If it is nevertheless extrapolated to larger
$t$, it disagrees with the measurement of
the combination $G_E^s +0.39 G_M^s$ at $t= -0.48$ GeV$^2$ by the
HAPPEX collaboration\cite{happex}.

As noted above, it is not clear whether the simple three pole 
approximation for $\FOS$ and $\FTS$ is justified. 
In the case of the isovector electromagnetic
form factors, {\it e.g.}, there is an important contribution from
the $\pi\pi$ continuum\cite{FrF60,HoP75a}. The left hand shoulder of the 
$\rho$ resonance is enhanced because of a singularity on the unphysical
sheet close to the physical region at $t_c=4\mpis-\mpis/M^2$.
The prime candidates for large continuum contributions to the
strange form factors are the $3\pi$ and $K\bar{K}$ states, which are the 
lightest intermediate states containing nonstrange and strange 
particles, respectively. However, it was shown that there
is no threshold enhancement in the $3\pi$ continuum\cite{BKM96}.
As a consequence, we focus on the $K\bar{K}$ continuum in the 
next section.

\subsection{$K\bar{K}$-Continuum}
\label{sec:kkbar}
In order to determine $K\bar{K}$ contribution to the spectral
functions in Eq.~(\ref{spectro}), we need the matrix elements
$\bra{N(p)}\bar{J}_N (0)\ket{K(k)\bar{K}(\bar{k})}\,v(\bar{p})$ and
$\bra{K(k)\bar{K}(\bar{k})}\bar{s}\gamma_\mu s\ket{0}$.

\subsubsection{Unitarity Bounds}
\label{sec:uni}
By expanding the $K\bar{K}\to N\bar{N}$ amplitude
in partial waves, we are able to impose the constraints of unitarity 
in a straightforward way. In doing so, we use the helicity
formalism\cite{JaW59}.
With $\lambda$ and $\bar{\lambda}$ being the nucleon and antinucleon
helicities, we write the corresponding $S$-matrix element as
\begin{eqnarray}
\label{kkbar:s_norm}
& &\bra{N(p,\lambda)\bar{N}(\pbar,\bar{\lambda})}{\hat S}\ket{K(k)\bar{K}
(\bar{k})}= \\ & &\hphantom{N(p,\lambda)}
 i(2\pi)^4\delta^4(p+\pbar-k-\bar{k})(2\pi)^2 \left[{64 t\over t-4 m_K^2}
\right]^{1/2} \bra{\theta,\phi,\lambda,\bar{\lambda}}{\hat S}(P)\ket{00}
\,,\nonumber
\end{eqnarray}
where $t=P^2=(p+\pbar)^2$ and $m_K$ is the kaon mass. The matrix element
$\bra{\theta, \phi, \lambda, \bar{\lambda}}{\hat S}(P)\ket{00}$
is then expanded in partial waves as\cite{JaW59,MHD97}
\begin{equation}
\label{kkbar:par_d}
S_{\lambda, \bar{\lambda}} \equiv
\bra{\theta, \phi, \lambda, \bar{\lambda}}{\hat S}(P)\ket{00}=\sum_J
\left({2J+1\over 4\pi}\right) b_J^{\lambda, \bar{\lambda}} \;
{\mathcal D}_{0\mu}^J(\phi, \theta, -\phi)^{\ast}\,,
\end{equation}
where ${\mathcal D}_{\nu\, \nu'}^J(\alpha, \beta, \gamma)$ is a
Wigner rotation matrix with $\mu=\lambda-\bar{\lambda}$.
The $b_J^{\lambda\, \bar{\lambda}}$ define partial waves of
angular momentum $J$. 
Only the $J=1$ partial waves contribute to the
spectral functions for vector currents. Moreover, because of 
parity invariance only two of the four partial 
waves for $J=1$ are independent. Choosing $\bpp$ and $\bpm$,
the unitarity of the $S$-matrix, $S^{\dag} S = 1$, requires that
\begin{equation}
\label{kkbar:ubs}
^4\!\!\sqrt{\frac{t-4M^2}{t-4\mks}}\left|b_1^{1/2, \pm 1/2}(t)\right|\leq 1 \,,
\end{equation}
for $t\geq 4\mns$.
Consequently, unitarity provides model-independent bounds on the
contribution of the physical region ($t\geq4\mns$) to the
imaginary part. To evaluate the DR, however, the $\bppm$ are
also needed in the unphysical region ($4\mks \leq t \leq 4\mns$),
where they are not bounded by unitarity.
Instead, we must rely upon an analytic continuation to
obtain the $\bppm$ in the unphysical region.

\subsubsection{Analytic Continuation}
\label{sec:ac}
A detailed discussion of
the analytic continuation procedure and its problems can
be found elsewhere\cite{HRM99a}, here we only give 
a brief overview.
To obtain the $b_1^{1/2,\pm 1/2}$  for $4\mks\leq t\leq 4\mns$, we
analytically continue physical amplitudes into the unphysical regime. 
However,
the analytic continuation (AC) of a finite set of experimental data 
with non-zero error is fraught with potential ambiguities.
Indeed, the AC is inherently unstable, and analyticity alone has no
predictive power. Additional information must be used in
order to stabilize the problem.

\begin{figure}[ht]
\centerline{\includegraphics*[width=8cm,angle=0]{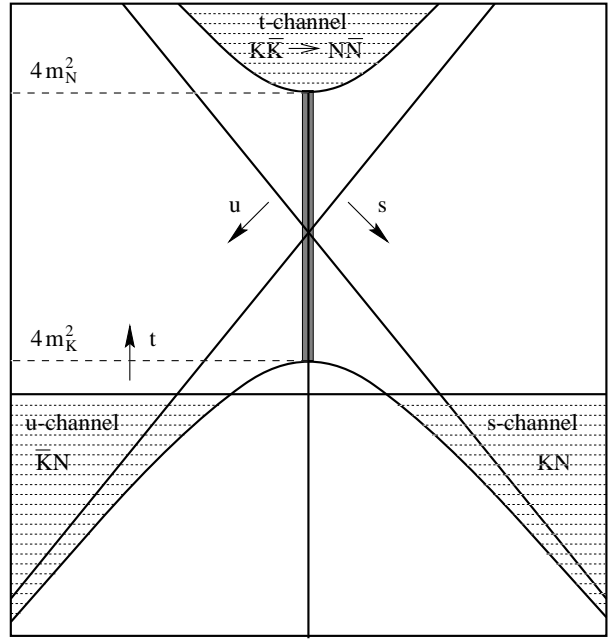}}
\vspace*{-.0in}
\caption{Mandelstam plane for $KN$ scattering. Physical regions are marked 
by dashed areas.}
\label{kkbar:fig1}
\end{figure} 
In order to illustrate these issues and the methods we adopt to
resolve them, we first  briefly review the kinematics of $KN$ scattering. 
It is useful to consider the $s$-, $u$-, and $t$-channel
reactions simultaneously,
\beqa
\nonumber
&\mbox{(a) $s$-channel:} & \qquad K (q_i) + N(p_i) \rightarrow K 
(q_f) +N(p_f) \,,\\
&\mbox{(b) $u$-channel:} & \qquad \bar{K} (-q_f) + N(p_i) \rightarrow
\bar{K} (-q_i) + N(p_f) \,,\nonumber \\
&\mbox{(c) $t$-channel:} & \qquad \bar{K} (-q_f) + K (q_i) \rightarrow
\bar{N}(-p_i) + N(p_f) \,,
\label{kkbar:chan}
\eeqa
where the four-momenta of the particles are given in parentheses.
In this notation the crossing relations between the different channels 
are immediately  transparent. The three processes can be described in 
terms of the usual Mandelstam variables
$s = (p_i+q_i)^2\,$, $u = (p_i-q_f)^2 \,$, and $t = (q_i-q_f)^2 \,$.
The ranges of $s,u,$ and $t$ in the
Mandelstam plane are shown in Fig \ref{kkbar:fig1}. 
The invariant amplitudes are defined in the whole plane and
simultaneously describe all three processes in the physical regions
marked by the dashed areas.
The physical values of the invariant amplitudes are obtained
when the Mandelstam variables are taken in the corresponding ranges.
In order to carry out the dispersion integrals, we require the 
$\bppm$ along the $t$-channel cut in the unphysical region. 
This is the gray shaded area in Fig.~\ref{kkbar:fig1} where the 
unitarity bounds do not apply.

\begin{figure}[ht]
\centerline{\includegraphics*[width=10cm,angle=0]{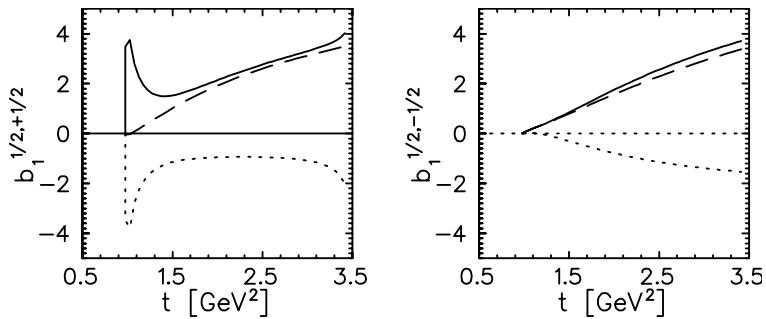}}
\vspace*{-.0in}
\caption{Analytically continued amplitudes $\bppm$. The dashed line
indicates the real part, the dotted line indicates the imaginary
part, and the solid line shows the absolute value.}
\label{kkbar:b1ppm}
\end{figure}

We begin with experimental $KN$ amplitudes in the s-channel region\cite{Hys92} 
and employ the method of backward DR to obtain the unphysical amplitudes
along the $t$-channel cut. In order to stabilize the 
numerical continuation, we subtract the rapidly varying nucleon pole 
terms which are known analytically.
We also divide out the asymptotic behavior of the amplitudes. We 
analytically continue the remainder using backwards DR's,
reinstate the asymptotic behavior, and add the pole terms afterwards.
The results\cite{HRM99a} for the amplitudes $\bppm$ are shown in 
Fig.~\ref{kkbar:b1ppm}.
The analytic continuation ist trustworthy up to $t\approx
2.0$ GeV$^2$. A clear resonance signature in $\bpp$, presumably
the $\phi$ resonance, is seen at the $K\bar{K}$ threshold.
\newpage

\subsubsection{Kaon Strangeness Form Factor}
\label{sec:ksf}
The second matrix element appearing in Eq.~(\ref{spectro}), 
$\bra{K(k)\bar{K}(\bar{k})} \bar{s}\gamma_\mu s \ket{0}$, is parametrized 
by the kaon strangeness form factor $\fks$:
\begin{eqnarray}
\bra{0}\bar{s}\gamma_\mu s\ket{K(k) \bar{K}(\bar{k})}&=&(k-\bar{k})_\mu
\fks(t)\,.
\end{eqnarray}
To obtain $\fks(t)$ we draw upon the known flavor content of the 
vector mesons. We use the flavor rotation arguments 
discussed in Sec.~\ref{sec:vmd} to obtain the strangeness form factor
from various parametrizations for the electromagnetic kaon form
factors\cite{MHD97,HRM99a}. 
This procedure should give a reliable estimate since 
the kaon has net strangeness.

The timelike kaon form factor is dominated by 
the $\phi(1020)$ resonance. We note that the flavor rotation applied
to the electromagnetic form factor only gives the relative size of 
the $\omega$ and $\phi$ contributions, but does not lead to the correct 
normalization $\fks(0)=-1$ which must be enforced by hand.
\begin{figure}[ht]
\centerline{\includegraphics*[width=8cm,angle=0]{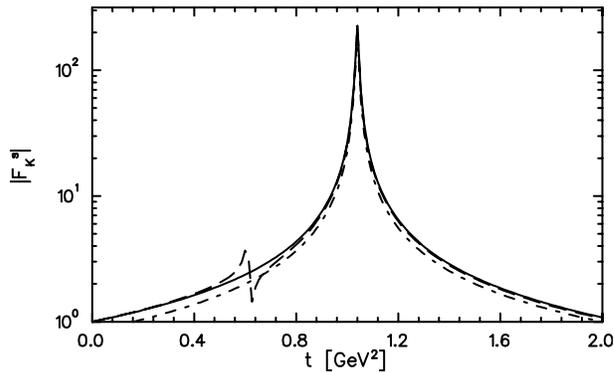}}
\vspace*{-.0in}
\caption{\label{drsd:fksfig}Different model parametrizations for $\fks$.
Full line shows simple VMD model with the $\phi$ only, dashed
line shows the flavor rotated VMD model including the $\omega$ and
$\phi$, and dash-dotted line shows Gounaris-Sakurai parametrization.
$K\bar{K}$ threshold is at $t\approx 1.0$ GeV$^2$.}
\end{figure}

In Fig.~\ref{drsd:fksfig} we plot $\fks$ extracted from three
different models for the electromagnetic form factor one of which
includes a contribution from the $\omega$.
We observe that all models reproduce the essential features of
$\fks$ as determined from $e^+e^-$ data and standard flavor rotation 
arguments. Since $\fks$ is needed for $t\geq 4\mks$, the 
$\omega$-contribution which gives rise to the bump around $t\approx 0.6 
\mbox{ GeV}^2$ in Fig.~\ref{drsd:fksfig} is irrelevant for our purpose.
When computing the leading strangeness moments, we find a less than 
10\% variation in the results when any of these different parametrizations 
for $\fks$ is used. 

\subsubsection{$K\bar{K}$ Spectral Function}
\label{sec:kkbarspec}
The exact relation of the spectral functions 
to the partial waves and the kaon strangeness form factor is\cite{MHD97}
\begin{eqnarray}
\label{drsd:imf1}
{\rm Im}\,  \FOS(t)&=&{\rm Re}\,\left\{
{\mn q_t\over 4 p_t^2}\left[
{E\over\sqrt{2}\mn}\bpm(t)-\bpp(t)\right]\fks(t)^{\ast}\right\}\,,\\
&& \nonumber \\
\label{drsd:imf2}
{\rm Im}\,  \FTS(t)&=&{\rm Re}\,\left\{
{\mn q_t\over 4 p_t^2}\left[
\bpp(t)-{\mn\over\sqrt{2}E}\bpm(t)\right]\fks(t)^{\ast}\right\}\,,
\end{eqnarray}
where
$p_t=\sqrt{t/4-\mns}\,$, $q_t=\sqrt{t/4-m_K^2}\,$, and $E=\sqrt{t}/2\,$.

On one hand, Eqs.~(\ref{drsd:imf1},\ref{drsd:imf2}) may be used to determine
the spectral functions from experimental data.
On the other hand, one can impose bounds on the imaginary parts
in the physical region by using Eq.~(\ref{kkbar:ubs}).
In order to obtain finite bounds for the Dirac and Pauli form factors
at the $N\bar{N}$ threshold, we build in the
correct threshold relation for the $\bppm$, $\bpm(t=4\mns)=\sqrt{2}
\bpp(t=4\mns)$. This is necessary to cancel the $1/p_t^2$
factor in Eqs. (\ref{drsd:imf1},\ref{drsd:imf2}). Strictly speaking,
the relation holds only for $t=4\mns$. For simplicity, however,
we assume this relation to be valid for all momentum transfers.
Consequently, we have\cite{MHD97}
\begin{eqnarray}
\label{kkbar:ubf1th}
|{\rm Im}\, \FOS(t)| &\leq& \frac{q_t^{3/2}}{2\sqrt{2p_t} 
( \sqrt{t} +2 \mn)}|\fks(t)|
\, ,\\
\label{kkbar:ubf2th}
|{\rm Im}\, \FTS(t)| &\leq& \frac{2\mn}{\sqrt{t}}
\frac{ q_t^{3/2}}{2\sqrt{2p_t}(\sqrt{t} +2 \mn)}
|\fks(t)| \, .
\end{eqnarray}
Similar expressions can be obtained for the Sachs form factors.

\begin{figure}[ht]
\centerline{\includegraphics*[width=10cm,angle=0]{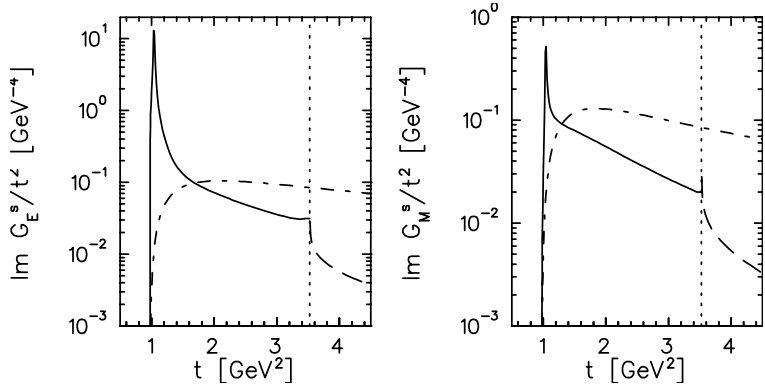}}
\vspace*{-.0in}
\caption{Spectral functions for the electric (left panel) and 
 magnetic radii (right panel). Solid line shows analytical continuation,
 dashed line shows unitarity bound, and dash-dotted line shows 1-loop
 result in the nonlinear $SU(3)$ sigma-model.}
\label{kkbar:specrad}
\end{figure}

In Fig.~\ref{kkbar:specrad}, we show the spectral function of the DR for 
the electric and magnetic radii as an example\cite{HRM99a}. We compare the 
analytical continuation (solid line) for $4\mks \leq t \leq 4\mns$
and the unitarity bound (dashed line) for $t \geq 4\mns$
to the 1-loop result in the nonlinear $SU(3)$ sigma-model
(dash-dotted line). Clearly
the perturbative 1-loop result misses the resonance structure at threshold.
Furthermore, it strongly overestimates the contribution in the physical 
region since rescattering contributions are neglected. These results
cast serious doubts on the reliability of any 1-loop model calculation of
the strange form factors.

In order to calculate the strange form factors, one needs the
contribution of all low-mass intermediate states in Eq.~(\ref{spectro}).
In principle it would be desirable to perform a full dispersion
calculation for all those states.
Since, at present, this is not possible due to the lack of
suitable experimental data, we combine the $K\bar{K}$ continuum
with the pole approximation for the remaining low-mass states in the
next section.

\subsection{Low-mass Contributions}
\label{sec:lowmass}
To make the connection to the leading strange moments,
we merge the pole approximation with the
exact $K\bar{K}$ continuum derived in the previous section\cite{HRM99b}.
The first step is to refit the isoscalar electromagnetic form factors
in order to see how much of the original $\phi$ pole strength is
provided by the $K\bar{K}$ continuum. We find that in $F_2$ all original
$\phi$ strength can be accounted for by the $K\bar{K}$ continuum, while
for $F_1$ some residual $\phi$ pole strength is required. 
This result justifies the identification of the pole around $t=1.0$ GeV$^2$
in electromagnetic analyses with the $\phi$\cite{Hoe76,MMD96}. 
The additional $\phi$ pole strength in $F_1$
could be due to a $\rho\pi$ resonance which
couples to the $\phi$. [Note that the branching ratio for the decay mode
$\phi\to\rho\pi$ is 12\%.]

In order to elucidate the consequences for the strange moments,
we obtain the $\omega$ and residual $\phi$ contributions to the 
strange form factors from a flavor rotation. The $K\bar{K}$ continuum
contribution which contains most of the $\phi$ strength is fixed
from Sec.~\ref{sec:kkbar}. However, the coupling of the $S'$
to strangeness is still unknown. In order to avoid the ambiguity 
from the asymptotic condition used previously\cite{Jaf89,HMD96s}, 
we allow a maximal coupling of the $S'$ to the strange current
and assess the uncertainty from varying the sign of this coupling.
For the leading strange moments, we define a {\it low-mass value}.
It contains the $\omega$, $\phi$, and  $K\bar{K}$ contributions which are
reasonably well under control. Adding the $S'$ contribution,
leads to a {\it reasonable range}, which quantifies the error
in the unknown coupling of the $S'$. Systematic errors from using
the pole approximation, however, can not be quantified this way.
The results\cite{HRM99b} for the leading moments are shown in Table~\ref{lmom}.
\begin{table}[htb]
\caption{}
\footnotesize{Table~\ref{lmom}.\ Leading strange moments 
from $K\bar{K}$ continuum plus $\omega$, $\phi$, and $S'$ poles.\vspace*{8pt}}
\centerline{
\begin{tabular}{|c||c|c|}
\hline Moment & {\it low-mass value} & {\it reasonable range} 
\\ \hline\hline
$\mu^s$ & $-0.28$ & $-0.15...-0.41$ \\ \hline
$\rosq$ & $0.42$ fm$^2$ & $(0.41...0.43)$ fm$^2$ \\ \hline
$\langle r^2 \rangle_M^s$ & $0.34$ fm$^2$ & $(0.32...0.36)$ fm$^2$ \\ \hline
\end{tabular}
\label{lmom}}
\end{table}
The magnetic moment seems to be most sensitive to the sign of the 
$S'$ coupling and varies by a factor of two, while the radii are 
relatively stable.
Compared to the pole analysis without the $K\bar{K}$ continuum,
the Dirac radius is somewhat larger. The magnetic moment 
can be smaller but still remains clearly negative.

\section{Connection to ChPT}
\label{sec:chpt}
\begin{figure}[ht]
\centerline{\includegraphics*[width=8cm,angle=0]{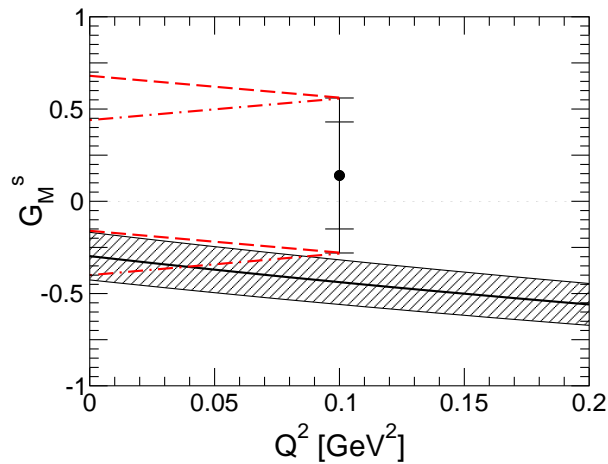}}
\vspace*{-.0in}
\caption{Error in extrapolation of $\GMS$ to $Q^2\equiv -t=0$. Circle
shows measurement of SAMPLE collaboration\protect\cite{sample}.
Dashed (dot-dashed) lines indicate extrapolation to $Q^2=0$
using $b_s^r= -1 (+1)$. Solid line gives low-mass DR
result, while shaded region indicates possible effects of higher mass 
states.}
\label{ChPTfig}
\end{figure}

In this section we illustrate how the dispersion results can be combined 
with chiral perturbation theory (ChPT). In ChPT, the strange moments 
$\mu^s$, $\rosq$, and $\langle r^2 \rangle_E^s$ depend to leading order
in the chiral expansion (LO) 
on unknown counterterms\cite{MuI97} and can at present
not be predicted. Hemmert \etal\cite{HMS98} have shown
that the strange magnetic radius $\rmsq$ is independent of unknown
counterterms at LO. Furthermore, for
the magnetic form factor, they derived a model independent 
relation between the strange and isoscalar electromagnetic form factor.
This leads to the LO prediction:
\beq
\rmsq = (-0.16 \ldots -0.60)\mbox{ fm}^2\,.
\label{op3}
\eeq
These numbers appear not to be compatible with the DR result from 
the previous section. They have been used to extrapolate the 
SAMPLE measurement of $G_M^s$ to $t=0$ in order to extract
the strange magnetic moment\cite{sample}.
However, at the next order in the chiral expansion there is
a new counterterm contribution to the magnetic radius. 
The coefficient $b_s^r$ of this counterterm is unknown.
In order to find out whether the inconsistency with DR's persists, 
we have extended the ChPT calculation to next-to-leading order 
(NLO)\cite{Ham02}. The result for the magnetic radius at NLO is
\begin{eqnarray}
\rmsq&=&-\left[0.04+0.3\,b_s^r(\mu=1.0\mbox{ GeV}^2)\right]\mbox{ fm}^2\,,
\label{chpt:nlo}
\end{eqnarray}
where $\mu$ is the renormalization scale.
The model independent contribution from LO
gets almost exactly cancelled by the loop contribution at
NLO. As a consequence, the strange magnetic radius at
NLO is dominated by the counterterm $b_r^s$.\footnote{This argument
is relatively insensitive to the value of $\mu$. Variation of $\mu$ 
between $m_\rho$ and $m_\Delta$ leads to a variation in the loop 
contribution in Eq.~(\ref{chpt:nlo}) of $\pm 0.04$ fm$^2$.}
We have determined $b_r^s$ from matching Eq.~(\ref{chpt:nlo}) to the DR
result and found $b_s^r \approx -1.1\,$,
which agrees with the expectation from dimensional analysis.
Consequently, the DR result and the chiral 
expansion are consistent to NLO. Furthermore,
due to the cancellation of the NLO and LO
loop contributions, the range for $\rmsq$ given in Eq.~(\ref{op3})
is not reliable. 
In Fig.~\ref{ChPTfig}, we illustrate the sensitivity of the extrapolation 
of the SAMPLE measurement of $\GMS$ to $Q^2\equiv -t=0$.
The dashed (dot-dashed) lines indicate extrapolation to $Q^2=0$
using $b_s^r= -1 (+1)$, respectively. For comparison,
the solid line gives low-mass DR result from Sec.~\ref{sec:lowmass},
while shaded region indicates possible effects of higher mass states.
Clearly, the error in the extrapolation to $Q^2=0$ is presently 
dominated by the statistical error of the experiment. 

\section{Conclusions}    
\label{sec:conclu}

Dispersion theory has provided many insights into the structure
of the nucleon's strange vector form factors. 
In particular, the role of the $K\bar{K}$ continuum to the
spectral function is now well understood.
Drawing upon rigorous bounds from unitarity and an analytic continuation
of $K\bar{K}\to N\bar{N}$ amplitudes, rescattering and resonance
contributions were shown to be crucial for the $K\bar{K}$ contribution 
to the spectral function\cite{MHD97,HRM99a}. 
This result casts doubt on any perturbative model calculation at
the 1-loop level where these effects are not included.
The apparent discrepancy in the magnetic radius between ChPT
at LO and the dispersion relation result disappears
at NLO where a new low-energy constant enters\cite{Ham02}. 

However, many open questions remain. Even though it is the most important
state,
the $K\bar{K}$ intermediate state is not the only relevant low-mass
state contributing to the spectral function. At present, a
full dispersion calculation for the other low-mass states 
(such as $3\pi, 5\pi,...,K\bar{K}\pi,K\bar{K}\pi\pi,...$) is not
feasible. Baryonic intermediate states are expected to be strongly
suppressed by unitarity. For the mesonic low-mass states,
one has to rely on the vector meson pole approximation. 
This introduces a systematic error that is difficult to quantify. 
In particular, there are some indications of a $\rho\pi$ resonance
in the $3\pi$ continuum, as well $K^*$ resonance effects in the
$K\bar{K}\pi$ and $K\bar{K}\pi\pi$ states. Finally, the large-$t$ behavior
from perturbative QCD might provide useful constraints on the 
sum over intermediate states in the spectral function.
More low-$t$ data on the nucleon's strange vector form factors will
be extremely useful in understanding these issues.

\section*{Acknowledgments}
This talk is based on work done in collaboration with D.~Drechsel,
U.-G.~Mei{\ss}ner, S.J.~Puglia, M.J.~Ramsey-Musolf, and
S.-L.~Zhu. This work was supported under NSF Grant No.\ PHY--0098645.

\end{document}